\documentclass[aps,prl,twocolumn,showpacs,superscriptaddress]{revtex4-1}
\usepackage{graphicx,amsmath,amssymb,amsfonts}
\usepackage[latin1]{inputenc}
\usepackage{array}
\usepackage{multirow}
\usepackage{float}
\usepackage{color}
\usepackage[normalem]{ulem}
\begin{document}

\title{Anisotropic magnetic properties and tunable conductivity in two-dimensional layered NaCrX$_2$ (X=Te,Se,S) single crystals}

\author{Jiale Huang}
\author{Bingxian Shi}
\author{Feihao Pan}
\author{Jinchen Wang}
\author{Juanjuan Liu}
\author{Daye Xu}
\author{Hongxia Zhang}
\author{Tianlong Xia}
\author{Peng Cheng}
\email[Corresponding author: ]{pcheng@ruc.edu.cn}
\affiliation{Laboratory for Neutron Scattering and Beijing Key Laboratory of Optoelectronic Functional Materials and MicroNano Devices, Department of Physics, Renmin University of China,
	Beijing 100872, China}

\begin{abstract}
Monolayer NaCrX$_2$ (X=Te,Se,S) were theoretically proposed to be two-dimensional intrinsic ferromagnetic semiconductors while their physical properties have not been thoroughly investigated in bulk single crystals. We report the single-crystal growth, structural, magnetic and electronic transport properties of NaCr(Te$_{1-x}$Se$_x$)$_2$ ($0\leqslant x \leqslant 1$) and NaCrS$_2$. For NaCr(Te$_{1-x}$Se$_x$)$_2$, the strong perpendicular magnetic anisotropy of NaCrTe$_2$ can be gradually tuned to be a nearly isotropic one by Se-doping. Meanwhile, a systematic change in the conductivity with increasing $x$ is observed, displaying a doping-induced metal-insulator-like transition. Under magnetic field larger than 30~kOe, both NaCrTe$_2$ and NaCrSe$_2$ can be polarized to a ferromagnetic state. While for NaCrS$_2$, robust antiferromagnetism is observed up to 70~kOe and two field-induced metamagnetic transitions are identified along H$\parallel$ab. These intriguing properties together with the potential to be exfoliated down to few-layer thickness make NaCrX$_2$ (X=Te,Se,S) promising for exploring spintronic applications.

\end{abstract}

\maketitle

\section{Introduction}
Magnetism in two dimensions has been a fascinating topic in condensed matter physics for decades. From the initial investigations on thin-film magnets to the recent discovery of two-dimensional (2D) magnetic order in ultra-thin van der Waals (vdW) materials, a wide range of possibilities for both spintronic applications and fundamental research have been opened up\cite{Burch2018,review1,reveiw2}. For few-layer vdW crystals, 2D magnetism has been realized in metallic Fe$_3$GeTe$_2$\cite{Fe3GeTe2_1,Fe3GeTe2_2}, semiconducting CrI$_{3}$/Gr$_{2}$Ge$_{2}$Te$_{6}$\cite{CrI3,Cr2Ge2Te6} and insulating FePS$_{3}$\cite{FePS3,FePS3_2} due to the large magnetic anisotropy which could counteract thermal fluctuations. These materials can serve as different building blocks of vdW heterostructures depending on their conductivity and exploring potential applications in novel magneto-electronic devices. Therefore magnetic anisotropy and conductivity are two key properties of 2D magnetic materials. Finding new 2D materials and ways to tune these properties would be quite important in the research of 2D magnetism.

Bulk CrTe$_2$ with 1$T$ phase is a vdW ferromagnet with in-plane magnetic anisotropy and Curie temperature of 310~K\cite{CrTe2_2015}. Remarkably, in untra-thin flakes or films, the easy axis of CrTe$_2$ changes from in-plane to out-of-plane and room-temperature ferromagnetism is retained\cite{Sun2020,CrTe2_film,CrTe2_CVD,CrTe2_PRM}. Moreover, a recent study identified a zigzag type antiferromagnetic order in monolayer CrTe$_2$\cite{CrTe2_AFM}, demonstrating the intricacy of 2D magnetism in this material. On the other hand, the intercalations of metal atoms into the vdW gap of CrX$_2$ (X=Te,Se,S) can form plenty of new phases. Although the structures of these intercalated CrTe$_2$ phases are non-vdW type, many of them are still able to be exfoliated into nanosheets while keep intriguing physical properties. For example, the room-temperature ferromagnetism in 10~nm-thick Cr-intercalated CrTe$_2$\cite{CrCrTe2} and the superionic behavior in 1.1~nm-thick AgCrS$_2$\cite{AgCrS2} have been reported recently. Especially for the latter, the structure of so-called AgCrS$_2$ monolayer consists of one Ag layer sandwiched between two [CrS$_2$] layers with actual formula AgCr$_2$S$_4$, which has been shown to be stable experimentally\cite{AgCrS2}.

The Na-intercalated NaCrX$_2$ (X=Te,Se,S) are recently proposed to be a class of monolayer ferromagnetic semiconductors by first-principles calculations\cite{NaCrTe_Calc}. Although the crystal structures are non-vdW type, their cleavage energies are comparable with other 2D materials. It also should be mentioned that, in previous research the vdW magnetic material CrTe$_2$ is mainly made by deintercalating the alkali metal of KCrTe$_2$. However the physical properties of NaCrTe$_2$ have not been thoroughly investigated until recently. Experimentally, NaCrTe$_2$ is determined to be an $A$-type antiferromagnet with T$_N$=110~K and perpendicular magnetic anisotropy\cite{NaCrTe2_IC,NaCrTe2_SC}. With applied magnetic field, spin-flip-driven giant negative and angle-dependent magnetoresistance are observed for single crystals of NaCrTe$_2$\cite{NaCrTe2_SC}. For NaCrSe$_2$ and NaCrS$_2$, although their antiferromagnetic transition temperature and structure have already been determined\cite{1973}, the anisotropic and field-induced magnetic properties are still unclear due to the lack of single crystals. Furthermore, chemical doping has been proved to be an effective way to tune the magnetic and transport properties of layered magnetic materials\cite{Fe3Co,PCheng_2020,PCheng_APL2020,NiFePS}. It would be interesting to explore the Se-doping effect on NaCrTe$_2$.

\begin{figure*}[htbp]
	\centering
	\includegraphics[width=\textwidth]{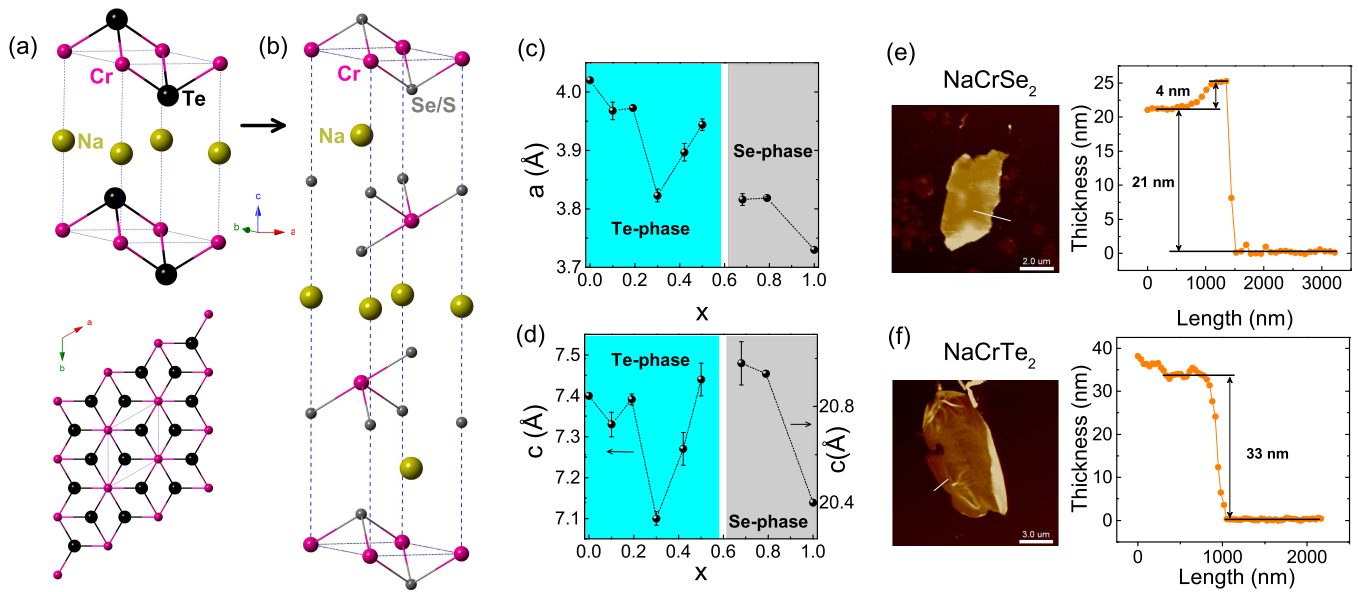}
	\caption {(a) Crystal structure of NaCrTe$_2$ (up) and top view (down) of CrTe$_2$ layer. The blue dotted lines represent the unit cell. (b) Crystal structure of NaCrSe$_2$ and NaCrS$_2$. (c,d) Lattice parameters obtained by fitting single crystal x-ray diffraction data for NaCr(Te$_{1-x}$Se$_x$)$_2$. (e,f) Atomic force microscopy images and height profile step of NaCrSe$_2$ and NaCrTe$_2$ nano-flakes mechanically-exfoliated onto a 300 nm SiO$_{2}$/Si substrate.} \label{Fig1}
\end{figure*}

In this paper, we report the successful growth of single crystals of NaCr(Te$_{1-x}$Se$_x$)$_2$ and NaCrS$_2$. These crystals show intriguing magnetic and transport properties, including field-induced spin-flip transitions, giant negative magnetoresistance, chemical doping controlled magnetic anisotropy and conductivity. Our findings suggest the NaCrX$_2$ series are promising candidates for further investigations in 2D limit.

\section{methods}

Single crystals of NaCr(Te$_{1-x}$Se$_x$)$_2$ and NaCrS$_2$ were grown by melting stoichiometric elements. High-purity Na, Cr, Te and Se/S were mixed in the mole ratio 1:1:2(1-$x$):2$x$. These reagents were mixed in alumina crucibles and sealed into an evacuated quartz tube. The assembly was heated up to $1050\,^{\circ}\mathrm{C}$ and maintained at this temperature for 24~h. Then it was slow-cooled to $800\,^{\circ}\mathrm{C}$ at a rate of $3\,^{\circ}\mathrm{C}$/h and annealed at this temperature for one day before furnace-cooled to room temperature. In order to make the reaction adequate, before heating to $1050\,^{\circ}\mathrm{C}$, the assembly would stay for ten hours at the temperatures slightly below the melting or boiling point of each reactant. The plane size of the obtained NaCr(Te$_{1-x}$Se$_x$)$_2$ single crystals is up to 6~mm$\times$6~mm, while the size of NaCrS$_2$ is smaller which is typically 1~mm$\times$1~mm. NaCrTe$_2$ is air-sensitive, its shinning surface could be oxidized and discolored after being put in the air for a few hours. With Se-doping, the crystal become less air-sensitive and it takes about two days for NaCrSe$_2$ to become degenerative in the air. NaCrS$_2$ is air-stable.

We characterized all samples with energy dispersive
x-ray spectroscopy (EDS, Oxford X-Max 50). For NaCr(Te$_{1-x}$Se$_x$)$_2$ with x=0.1 and x=0.2, the EDS value is quite close to the nominal value. For x$\textgreater$0.3, the doping concentration may slightly deviate from the nominal value (the estimated error is about 20\%). In order to be accurate, all measured samples have been carefully checked by EDS. The descriptions in this paper about doping level $x$ all refer to the EDS values.

X-ray diffraction (XRD) of the samples were collected from a Bruker D8 Advance X-ray diffractometer and a Bruker D8 VENTURE single-crystal diffractometer using Cu K$_{\alpha}$ radiation. Magnetization and electrical transport measurements were carried out in Quantum Design MPMS3 and PPMS-14T, respectively. The dimensions of exfoliated NaCrTe$_2$ and NaCrSe$_2$ nanoflakes were checked by a Bruker edge dimension atomic force microscope. 

\begin{figure*}[htbp]
	\centering
	\includegraphics[width=\textwidth]{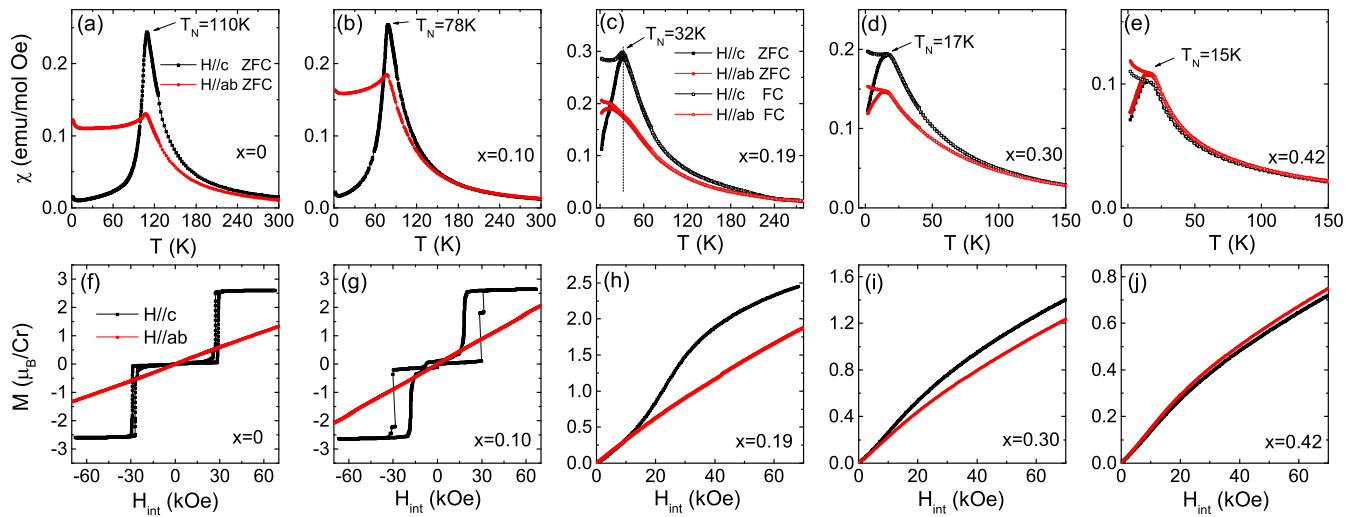}
	\caption {(a)-(e) The temperature dependent magnetic susceptibilities measured on NaCr(Te$_{1-x}$Se$_x$)$_2$ ($0\leqslant x \leqslant 0.42$) under magnetic field applied along $ab$-plane or $c$-axis. (f)-(j) Magnetization isotherms measured on the same crystals at $T$=2~K. For $x$=0 and $x$=0.10, the hysteresis loops are presented.} \label{Fig2}
\end{figure*}

\section{Results and discussions}

\subsection{Crystal structure of NaCrX$_2$}

As shown in Fig. 1(a) and confirmed by XRD analysis, NaCrX$_2$ crystallizes in a hexagonal structure with the space group of $P$-3$m$1 for X=Te and $R$-3$m$ for X=Se/S, same as previous reports\cite{1973,NaCrTe2_IC,NaCrTe2_SC}. The crystal structure of NaCrTe$_2$ can be considered as the intercalation of Na atoms between 1$T$-CrTe$_2$ layers. Na and Cr atoms stack alternately along the $c$-axis in the same site. The Cr atoms form a triangular lattice in the $ab$-plane. When Te was substituted by Se or S, the major change in crystal structure is the interlayer stacking order. Comparing with NaCrTe$_2$, for NaCrSe$_2$ and NaCrS$_2$, both Na- and Cr-triangular lattice layers are stacked along the $c$-axis with (1/3 1/3) translation in the $ab$-plane. So the $c$-lattice parameter has almost tripled. These two different structures are referred as 'Te-phase' and 'Se-phase' respectively in the following descriptions. 

For NaCr(Te$_{1-x}$Se$_x$)$_2$ ($0\leqslant x \leqslant1$), the lattice parameters were obtained by refining the single crystal x-ray diffraction data and plotted in Fig. 1(c) and (d). The results show that the samples with $x \leqslant 0.5$ maintain the Te-phase while the samples with $x \geqslant 0.68$ are confirmed to have the Se-phase. Therefore the phase boundary may exist near $x$=0.6, although it has not been accurately determined. The precession images from the single crystal x-ray diffraction data display streaking features for most doped samples, especially for the heavily doped ones. This indicates that notable stacking disorders and possible phase separations may exist in NaCr(Te$_{1-x}$Se$_x$)$_2$ (Figure S2 in Supplementary Materials\cite{Supple}). In the Te-phase zone, both the $a$- and $c$-lattice parameters do not follow a monotonic change with increasing $x$. Compare with NaCrTe$_2$, there is a 7\% shrinkage for the $a$-axis of NaCrSe$_2$. The $c$-lattice parameter for NaCrS$_2$ is 19.485$\AA$ , which decreases about 4\% compared with that of NaCrSe$_2$. 

Previous calculations of the cleavage energies suggest that NaCrX$_2$ can be exfoliated to a thickness of a few layers\cite{NaCrTe_Calc}. 
We performed the mechanical exfoliation of bulk NaCrX$_2$ single crystals using Scotch tape. Nanosheets of NaCrTe$_2$ and NaCrSe$_2$ with thickness 20-30~nm could be obtained, which is demonstrated by the atomic force microscopy images in Fig. 1(e) and (f). Recently Jing Peng $et~al.$ demonstrated that isostructural AgCrS$_2$ can be exfoliated into 1.1~nm nanosheet which consists of one Ag-layer sandwiched between two CrS$_2$-layers\cite{AgCrS2}. Therefore this class of materials are promising 2D materials for further investigations. Future investigation on whether NaCrX$_2$ is stable under similar exfoliation method as AgCrS$_2$ could be stimulated.

\subsection{Magnetic properties of NaCr(Te$_{1-x}$Se$_x$)$_2$}

The temperature dependent magnetic susceptibility $\chi$(T) and isothermal magnetization M(H) of NaCr(Te$_{1-x}$Se$_x$)$_2$ single crystals are shown in Fig. 2. Demagnetization corrections with methods used in our previous publication have been applied on the $H \parallel c$ data and the applied magnetic field
H$_{app}$ in $M(H)$ curve is replaced by the internal
field H$_{int}$\cite{PCheng_APL2020}. For NaCrTe$_2$, $\chi$(T) curve under H$\parallel$c exhibits a sharp drop down to nearly zero below T$_N$=110~K in contrast to the weak cusp and plateau-like feature under H$\parallel$ab. This suggests the development of an $A$-type antiferromagentic order (ferromagnetic intralayer and antiferromagnetic interlayer couplings) with moment aligned along $c$-axis. The Curie-Weiss fit on the high-temperature paramagnetic susceptibility reveals 
$\mu_{eff}/Cr = 3.8~\mu_{B}$ and $\theta_{CW} = 155~K$. The large positive $\theta_{CW}$ value indicates the strong intralayer ferromagnetic correlations. The hysteresis loops at 2~K under H$\parallel$c and H$\parallel$ab demonstrate the strong perpendicular magnetic anisotropy (PMA) and a spin-flip transition to the ferromagnetic state near H$_{ab}$=30~kOe. These observations are similar as previous report\cite{NaCrTe2_SC}. The saturation moment for NaCrTe$_2$ is $2.6~\mu_{B}/Cr$, which is a bit lower than the theoretical value $3.0~\mu_{B}$ for Cr$^{3+}$ in a localized model.

For x=0.1, T$_N$ decreases to 78~K while the features of $A$-type antiferromagnetic order and perpendicular magnetic anisotropy still persist. The spin-flip transition near 30~kOe also exists but with a much larger hysteresis comparing with that of $x$=0. An important modification of magnetic anisotropy is that the PMA seems to get weakened with Se-doping. If we choose the ratio M$_c$/M$_{ab}$ at H=60~kOe and T=2~K as a criterion, it decreases from 2.3 for $x$=0 to 1.5 for $x$=0.10. 

With increasing doping concentration $x$, T$_N$ continuously shifts to lower temperature and the spin-flip transition gradually vanishes as shown in Fig. 2(c)-(d) and (h)-(j). The M$_c$/M$_{ab}$ at 60~kOe and 2~K also gradually decreases from 1.4 ($x$=0.19) to 1.2 ($x$=0.30), and finally to 0.95 ($x$=0.42). This means the magnetic anisotropy gradually evolves from PMA to a slightly preferred in-plane magnetization. On the other hand, for $x \geqslant 0.19$ , there is a bifurcation between zero-field-cooling (ZFC) and field-cooling (FC) magnetization below T$_N$, which implies the emergence of a spin-glass state. In addition, the Curie-Weiss fit on the doped samples reveals similar values of effective moment but lower $\theta_{CW}$ values (131~K-106~K), suggest a slightly weakened ferromagnetic correlations.

\begin{figure}
	\includegraphics[width=7.5cm]{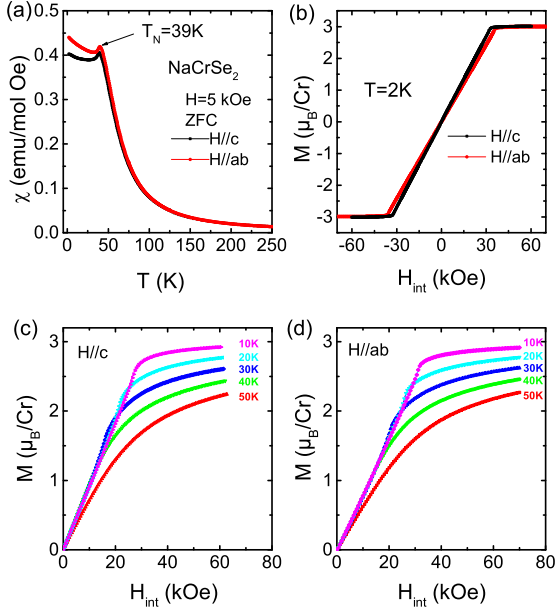}
	\caption {Anisotropic magnetization data for NaCrSe$_2$ single crystal: (a) The temperature dependent magnetic susceptibilities at $H$=5~kOe. (b) Magnetic hysteresis loops at $T$=2~K. (c) and (d) Magnetization isotherms measured at selected temperatures under $H \parallel c$ and $H \parallel ab$, respectively. } \label{Fig3}
\end{figure}

\begin{figure}
	\includegraphics[width=7cm]{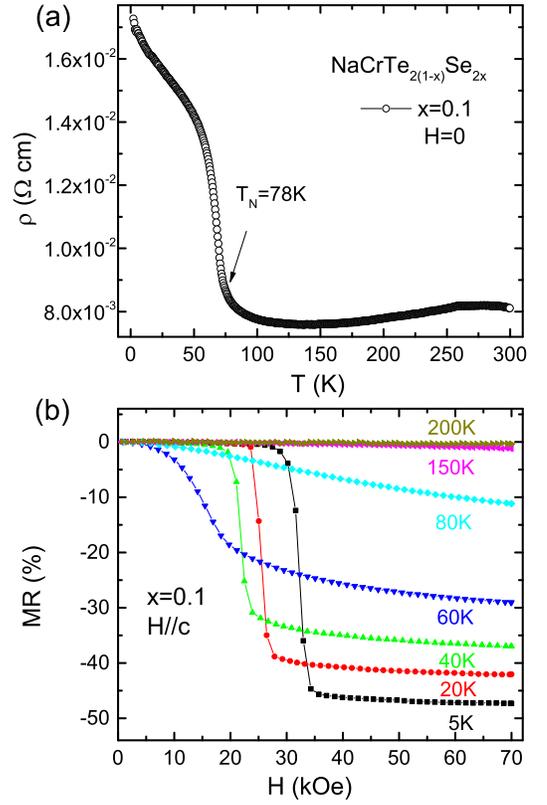}
	\caption {(a) The temperature dependent resistivity of  NaCr(Te$_{1-x}$Se$_x$)$_2$ with $x$=0.1 under zero field. (b) Isothermal magnetoresistance (MR) of $x$=0.1 under $H\parallel c$. } \label{Fig4}
\end{figure}

The samples with $0\leqslant x \leqslant 0.42$ discussed above all belongs to the Te-phase. Our XRD analysis reveal that at least from $x$=0.68, NaCr(Te$_{1-x}$Se$_x$)$_2$ enters the Se-phase. Let us first discuss the magnetic properties of NaCrSe$_2$ which is shown in Fig. 3. A cusp at T$_N$=39~K is observed in the $\chi(T)$ curve indicating an antiferromagnetic transition. On the other hand, fitting the high-temperature paramagnetic data to the Curie-Weiss law yields 
$\mu_{eff}/Cr = 3.8~\mu_{B}$ and $\theta_{CW} = 108~K$. The positive $\theta_{CW}$ temperature indicates ferromagnetic correlations are still strong in each individual layer, while the interlayer coupling is antiferromagnetic, similar as NaCrTe$_2$ and many other 2D layered magnetic materials.

\begin{figure*}[htbp]
	\centering
	\includegraphics[width=\textwidth]{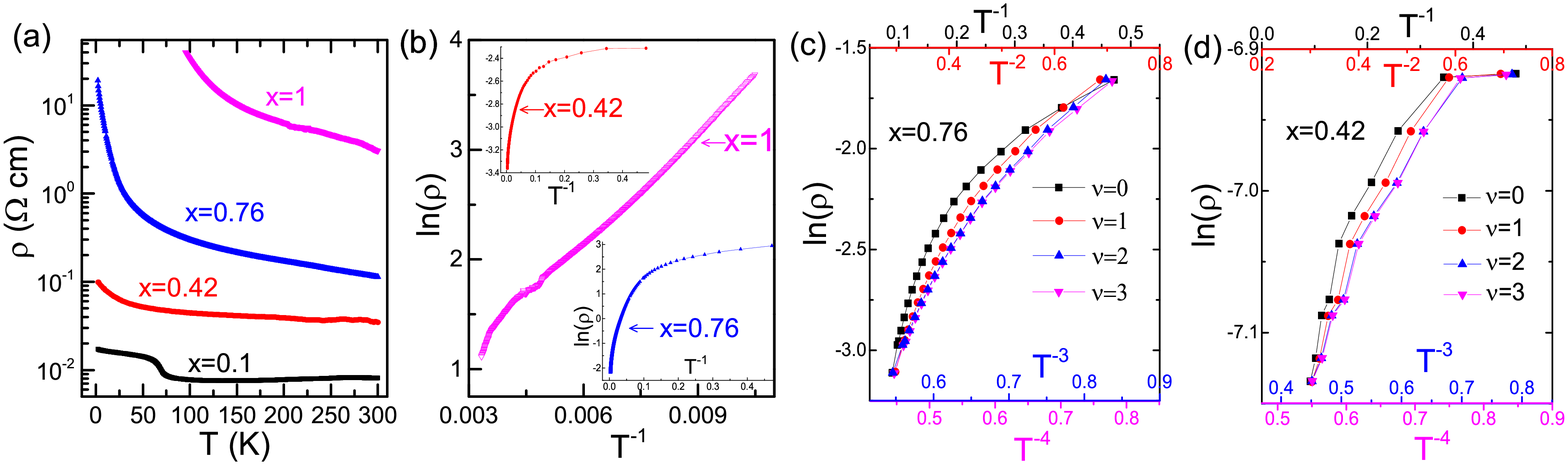}
	\caption {(a) Temperature dependent resistivity data for NaCr(Te$_{1-x}$Se$_x$)$_2$ ($H$=0). (b) Ln$\rho$ versus T$^{-1}$ plot of $x$=1 for all measured temperatures. The inset shows same plots for $x$=0.42 and $x$=0.76. (c,d) Low temperature Ln$\rho$ versus T$^{-1/(\nu+1)}$ plots of $x$=0.76 and $x$=0.42, respectively. $\nu$ is the parameter from VRH formula.} \label{Fig5}
\end{figure*}

In contrast to the strong anisotropic magnetization under H$\parallel$c and H$\parallel$ab for NaCrTe$_2$, the magnetic anisotropy of NaCrSe$_2$ is much smaller. Without applying a demagnetization correction, NaCrSe$_2$ appears to have essentially zero anisotropy. After the correction, a weak PMA could be identified from the M(H) curves at 2~K [Fig. 3(b)]. Under $H \geqslant 30~kOe$ for both directions and $T$=2~K, the magnetization becomes saturated with a saturation moment of $3.0~\mu_{B}/Cr$, which is larger than that of NaCrTe$_2$ and accurately equals to the theoretical value of Cr$^{3+}$ in a localized model. It should be noted that, before reaching saturated value, the magnetization increases quickly and linearly with increasing field along both directions. This indicate that although the spins of NaCrSe$_2$ are antiferromagnetically aligned at low field, they could be continuously canted along the field direction with increasing field. Fig. 3(c) and (d) show the M(H) curves at different temperatures. The saturation moment gradually decreases with increasing temperature. At 50~K which is well above T$_N$, the M(H) curve still exhibits a nonlinear curvature and a large moment ($\sim 2.2~\mu_{B}/Cr$) at 60~kOe. In addition, the M(T) curve also deviates from Curie-Weiss behavior below 200~K, which are actually observed for all NaCrX$_2$ (X=Te,Se,S) samples. These observations suggests that strong magnetic fluctuations or short-range magnetic order may develop well above T$_N$.

For $x$=0.68 and $x$=0.79 with Se-phase, the spin-glass behavior are quite similar as $x$=0.42 with Te-phase. The related magnetization data is not shown here but the results are plotted in the phase diagram of Fig. 6. Further discussions about the evolution of anisotropic magnetic properties of NaCr(Te$_{1-x}$Se$_x$)$_2$ will be presented in the following section with the phase diagram.

\subsection{Tunable conductivity in NaCr(Te$_{1-x}$Se$_x$)$_2$}

Previously, NaCrTe$_2$ was reported to exhibit a metal-insulator-like transition due to the formation of $A$-type antiferromagnetic ordering\cite{NaCrTe2_SC}. Similarly, the resistivity of $x$=0.1 sample has a typical metallic behavior above 130~K [Fig .4(a)], while a sharp jump appears at T$_N$=78~K with deceasing temperature which should result from the localization of charge carriers by long-range antiferromagnetic order. With further deceasing temperature, the resistivity continues to increase with a smaller slope which might be attributed to the impurity scattering effect by chemical-doping. Under magnetic field along $c$-axis, a giant negative magnetoresistance appears with maximal value up to 48\% at 5~K and 70~kOe. This behavior is directly associated with the field-induced spin-flip transition and due to the reduced spin scattering to the electrons in the ferromagnetic state comparing with that in the antiferromagnetic states.

Comparing with the previously reported resistivity of NaCrTe$_2$\cite{NaCrTe2_SC}, we noticed that the value for $x$=0.1 approximately increases by an order of magnitude. As shown in Fig. 5(a), intriguingly, with further increasing Se-doping concentration $x$, the temperature dependent resistivity of NaCr(Te$_{1-x}$Se$_x$)$_2$ becomes fully semiconducting-like and the absolute value of resistivity continuously increases by several orders. For NaCrSe$_2$ with $x$=1, the resistivity is about $\sim$ 10~$\Omega.cm$ and the data below 95~K cannot be obtained due to the upper limit of PPMS measurement. Thus a Se-doping induced metal-insulator transition is observed for NaCr(Te$_{1-x}$Se$_x$)$_2$.

Density functional theory calculations have shown that both NaCrTe$_2$ and NaCrSe$_2$ are semiconductors, their band gaps are 0.59~eV and 0.77~eV respectively\cite{NaCrTe_Calc}. In order to check whether the doping controlled conductivity is due to the gradual increasing of the band gap, we have tried to fit the resistivity data using the thermal activation model as described by $\rho$=$\rho_0$exp(E$_a$/2k$_B$T). Where $E_a$ is the energy gap and $k_B$ is the Boltzmann constant. Therefore a fine fitting result using this formula can only be obtained when ln$\rho$ and T$^{-1}$ follows a linear relationship. However, as shown in Fig. 5(b), there are not well defined linear relations between ln$\rho$ and T$^{-1}$ for all samples, especially for $x$=0.42 and $x$=0.76.   

\begin{figure}
	\includegraphics[width=7cm]{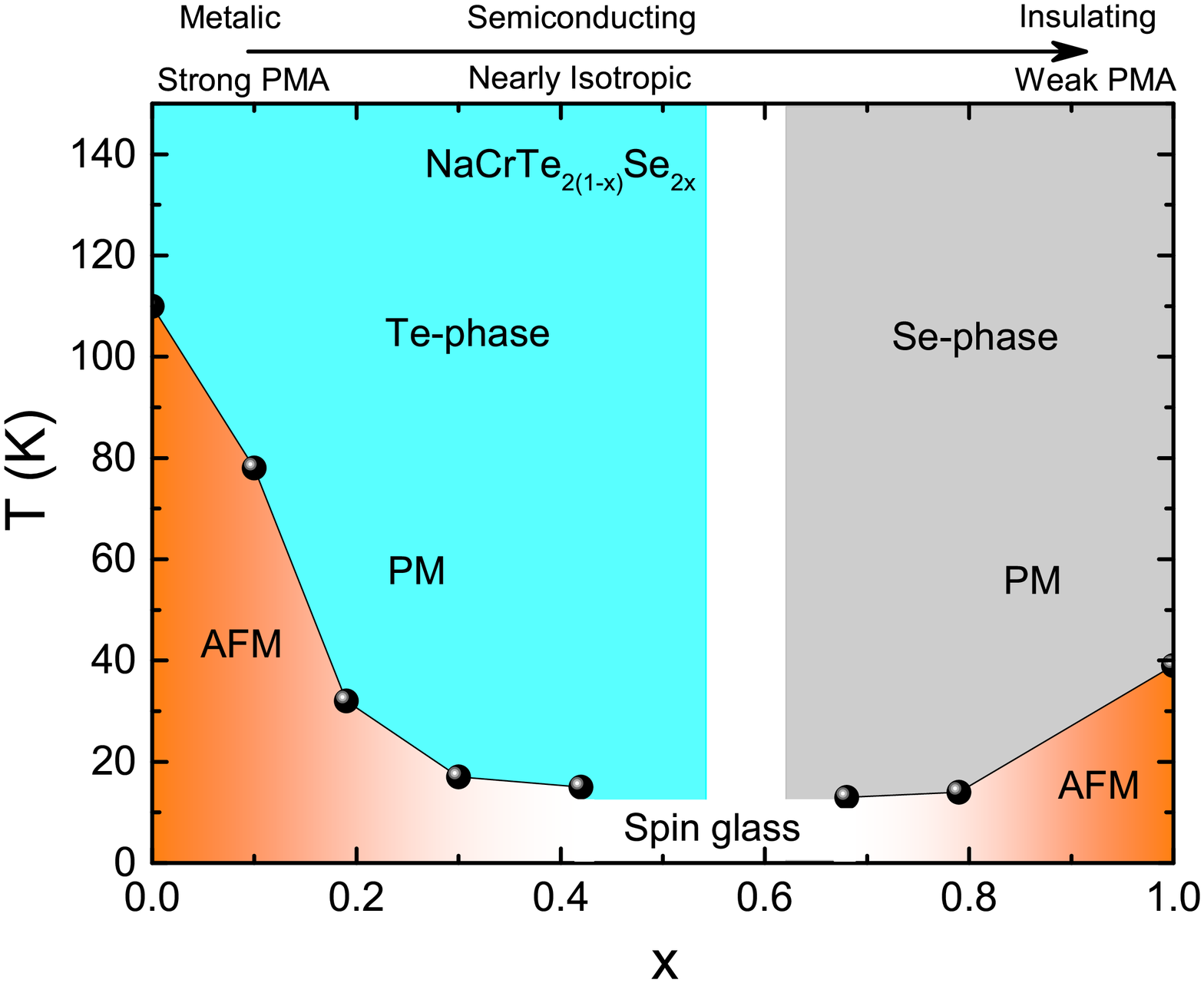}
	\caption {Temprature versus doping phase diagram for NaCr(Te$_{1-x}$Se$_x$)$_2$ ($0\leqslant x \leqslant 1$).} \label{Fig6}
\end{figure}

For chemical doped samples, the doping-induced strong disorder potential might trap itinerant electrons and lead to a metal-insulator transition, which is the famous Anderson localization\cite{Anderson}. For an Anderson insulator at low temperatures, there are electronic states trapped in the vicinity of the Fermi surface and the hopping transport by localized electrons would be described by the variable range hopping (VRH) model $\rho$=$\rho_0$exp(T$_0$/T)$^{1/(\nu+1)}$\cite{LiFeSe}. In this formula, $\nu$=0 denotes the traditional insulator with band gap (same as thermal activation model described above). $\nu$=1, 2 and 3 correspond to one-, two-, and three-dimensional material with Anderson localization, respectively. In Fig. 5(c) and (d), the ln$\rho$ versus T$^{-1/(\nu+1)}$ plots for $x$=0.42 and 0.76 from 2~K to 10~K are presented. Although a perfect straight line is not observed, the formula using $\nu$=2 or $\nu$=3 would clearly give a much better fitting results comparing with that using $\nu$=0, especially for $x$=0.76. Our results indicate that the significant enhancement of resistivity in NaCr(Te$_{1-x}$Se$_x$)$_2$ may partially originate from the Anderson localization. Besides, the tuning of band gap by Se-doping would also be possible. The electronic transport properties of NaCr(Te$_{1-x}$Se$_x$)$_2$ may contains contributions from these two aspects.

\begin{figure*}[htbp]
	\centering
	\includegraphics[width=\textwidth]{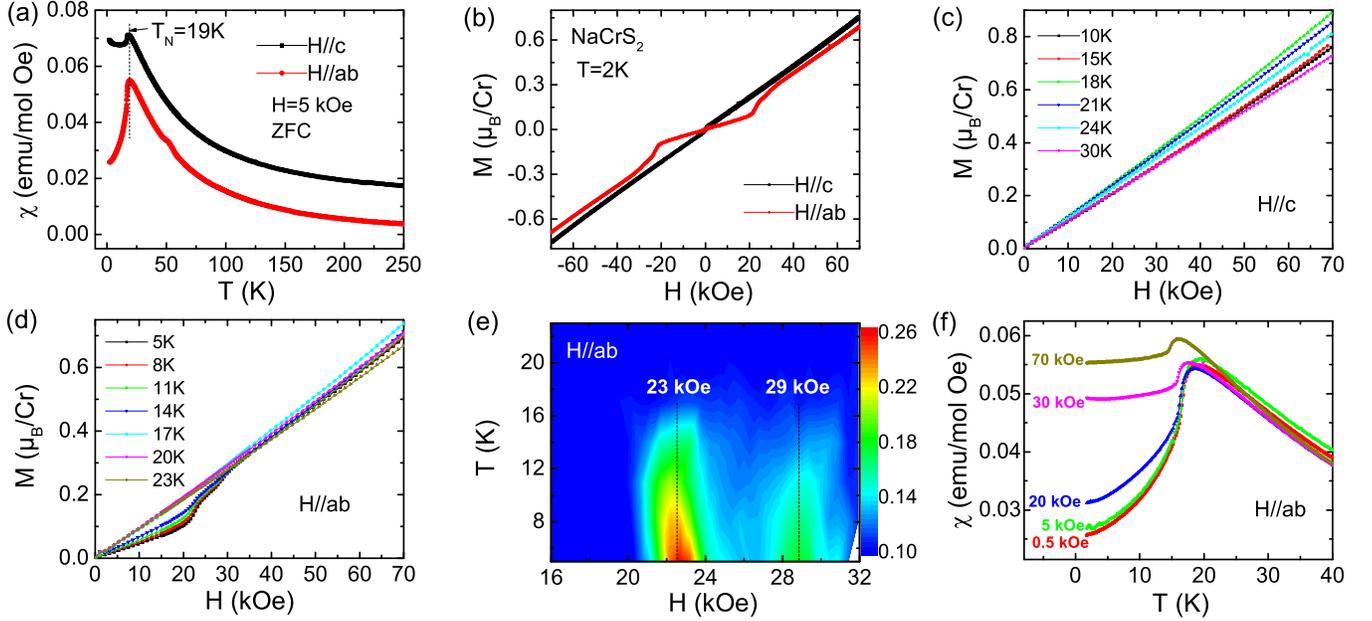}
	\caption {Anisotropic magnetization data for NaCrS$_2$ single crystal: (a) The temperature dependent magnetic susceptibilities at $H$=5~kOe. (b) Magnetic hysteresis loops at $T$=2~K. (c) and (d) Magnetization isotherms measured at selected temperatures under $H \parallel c$ and $H \parallel ab$, respectively. (e) Contour plot of $dM/dH$ as a function of temperature and field along $H \parallel ab$. (f) The temperature dependent magnetic susceptibilities under different applied field near the antiferromagnetic transition. } \label{Fig7}
\end{figure*}

\subsection{Discussion on the phase diagram of NaCr(Te$_{1-x}$Se$_x$)$_2$}

The $T$-$x$ phase diagram of NaCr(Te$_{1-x}$Se$_x$)$_2$ is presented in Fig. 6, which summarizes the major experimental results above. Three features should be mentioned about this phase diagram. Firstly, considering either from Se-doped NaCrTe$_2$ or Te-doped NaCrSe$_2$, the antiferromagnetic transition temperature gradually decreases and the system enters a spin-glass state with increasing doping concentration. Secondly, the magnetic anisotropy can be effectively tuned by chemical doping in this system. NaCrTe$_2$ possesses a strong PMA and it continuously gets weakened and evolves into a nearly isotropic magnetic behavior at $x$=0.42 while maintain the Te-phase. For NaCrSe$_2$ at the other end, a weak PMA is identified. Thirdly, A systematic change in the conductivity of NaCr(Te$_{1-x}$Se$_x$)$_2$ occurs on increasing $x$, from bad metallic behavior in NaCrTe$_2$ to the semiconducting or insulating behavior in NaCrSe$_2$. 

The doping-induced disorder effect and stacking faults should be responsible for the emergence of the spin glass state\cite{SG1}. It also may play an important role in tuning the magnetic anisotropy in NaCr(Te$_{1-x}$Se$_x$)$_2$. In previous investigations on materials for high-density magnetic recording media, chemical disorder has been shown to have important influence on the magnetocrystalline anisotropy energy (MAE)\cite{FePt,FeNi,FeCo}. It may either drastically reduce the MAE or tune the MAE to a maximum value\cite{FePt,FeNi,FeCo}. For vdW material, recent studies reveal that Ni$_{1-x}$Fe$_x$PS$_3$ and Fe$_{5(1-x)}$Co$_{5x}$GeTe$_2$ enable chemical tuning of easy-plane and easy-axis anisotropies\cite{NiFePS,PCheng_APL2020,May}. Magnetic anisotropy is a key property of 2D magnets, which is required for counteracting thermal fluctuations. There have been reports about pressure control of magnetic anisotropy on CrGeTe$_6$\cite{Pressure} and tensile strain-tunable magnetic anisotropy in monolayer CrX$_3$ (x=Cl, Br, I)\cite{Strain}. Our study provides a new example for chemical disorder or stacking fault controlled magnetic anisotropy in 2D magnetic material NaCr(Te$_{1-x}$Se$_x$)$_2$.

In addition, one should notice that the magnetic anisotropy also makes a big difference between NaCrTe$_2$ and NaCrSe$_2$, both of them seem to have no evident chemical disorder. A possible speculation is that, for NaCrSe$_2$ with relatively much lighter Se element, the spin-orbit coupling effect may become weak and results in a weak PMA\cite{MAE}. As we mentioned above, the saturation moment of NaCrSe$_2$ agrees well with the expectation of Cr$^{3+}$ spin $S$=3/2 model without orbital moment while that of NaCrTe$_2$ has a notable smaller value which might be due to the enhanced spin-orbital coupling. 

The tuning of conductivity by chemical doping in 2D magnetic material has been rarely reported. This effect found in NaCr(Te$_{1-x}$Se$_x$)$_2$, possibly due to Anderson localization and the change of band structure, may have important applications in designing novel spintronic devices. Particularly, the chemical doping in NaCr(Te$_{1-x}$Se$_x$)$_2$ could simultaneously tune both the conductivity and magnetic anisotropy. 

. 

\subsection{Field-induced metamagnetic transitions in NaCrS$_2$}

For the magnetic properties of NaCrS$_2$, so far as we know, there has not been any investigations on the single crystals. We present the anisotropic magnetization data on NaCrS$_2$ single crystals in Fig. 7. The different temperature dependent features of $\chi_{ab}$ and $\chi_{c}$ below T$_N$=19~K indicate the ordered magnetic moment should be confined within $ab$-plane [Fig. 7(a)]. Fig. 7(b) present the magnetic hysteresis loops at 2~K. Under magnetic field applied along the hard-axis ($H \parallel c$), a linear relationship between magnetization and field is observed. But for field applied along the easy $ab$-plane, a sudden magnetization jump is revealed at around $25~kOe$ suggest the occurrence of field-induced metamagnetic transition. From the magnetization isotherms at higher temperatures [Fig. 7 (c) and (d)], the linear behavior persists for $H \parallel c$ and the magnetization jump gradually weakens with increasing temperature for $H \parallel ab$. The contour plot of $dM_{ab}/dH$ in Fig. 7(e) reveals that there are actually two metamagnetic transitions which appear at $H$=23 kOe and $H$=29 kOe respectively. These two transitions gradually disappear near T$_N$. The transition at lower field slightly softens with increasing temperature similar as that for a conventional spin-flop transition. 

Early neutron diffraction result on NaCrS$_2$ powders have determined its magnetic structure to be a in-plane helical one\cite{1973}. The spins of Cr have antiferromagnetic coupling along the $c$-axis, while in the $ab$-plane they are aligned in a helimagnetic order with rotating angle $\phi=33^{\circ}$ in adjacent (110) planes\cite{1973}. This ground state magnetic structure is consistent with the anisotropic magnetic behavior of our single-crystal sample. It is interesting to point out that the Curie-Weiss fit gives still positive $\theta_{CW}$ values (18~K for $H \parallel ab$ and 31~K for $H \parallel c$) which suggests the ferromagnetic correlations still exist. In addition, we noticed that the paramagnetic susceptibility and corresponding fitted $\theta_{CW}$ values have large anisotropy. Whether this really means the anisotropy on the exchange interactions need further researches to clarify, as this kind of difference can arise solely from the
single-ion anisotropy even without exchange interaction as in a previous theoretical research\cite{TTT}.

Furthermore, the observed field-induced metamagnetic transitions should have significant influences on this complex helimagnetic order. Firstly, the magnetic order under field above $30~kOe$ should also be antiferromagnetic-like. This is supported by the cusp-like feature in the $\chi_{ab}-T$ curve at T$_N$, as shown in Fig. 7 (f). The antiferromagnetic transition seems to be quite robust under field. T$_N$ only has a slight shift to 16~K under $H=70~kOe$. In addition, M(H) curve follows a linear relationship up to 70~kOe with a small value of moment ($\sim$ 0.7$\mu_B$/Cr) under this field. Secondly, for $H_{ab} \leqslant 20~kOe$ the magnetic susceptibilities have a sharp drop below T$_N$. In contrast, for $H_{ab}=30~kOe$ and 70~kOe, the magnetic susceptibilities exhibit a plateau below T$_N$, which is a typical feature of canted antiferromagnetic magnetic order. Therefore, new types of antiferromagnetic order are expected under in-plane magnetic field, which would be an interesting topic for further investigations using neutron scattering.        

\section{Conclusions}

In summary, the physical properties of NaCrX$_2$ (X=Te,Se,S) are investigated in single-crystal form. For NaCr(Te$_{1-x}$Se$_x$)$_2$, a field-induced spin-flip transition together with a giant negative magnetoresistance are observed at $x\leqslant$0.1. At higher doping level, a spin glass state emerges. The most prominent feature is that both the magnetic anisotropy and the conductivity can be effectively tuned by $x$. For NaCrS$_2$, two magnetic field induced metamagnetic transitions are identified. We further demonstrate that these crystals can be mechanically-exfoliated into nano-flakes. These properties would make NaCrX$_2$ a promising material playground for further investigations on 2D magnetism and designing novel magneto-electronic devices. Furthermore, our samples also provide a route to vdW layered Cr$X_2$ through deintercalation.

\section*{Acknowledgement}
This work was supported by the National Natural Science Foundation of China  (No. 12074426, No. 11227906 and No. 12004426), the Fundamental Research Funds for the Central Universities, and the Research Funds of Renmin University of China (Grants No. 22XNKJ40).

\bibliography{NCT}{}
\end{document}